\documentclass[authoryear]{article}
\usepackage[preprint]{neurips_2020}
\usepackage[utf8]{inputenc} % allow utf-8 input
\usepackage[T1]{fontenc}    % use 8-bit T1 fonts
\usepackage{url}            % simple URL typesetting
\usepackage{booktabs}       % professional-quality tables
\usepackage{amsfonts}       % blackboard math symbols
\usepackage{nicefrac}       % compact symbols for 1/2, etc.
\usepackage{microtype}      % microtypography
\usepackage{amsmath}
\usepackage{multirow}
\usepackage{graphicx} % for \scalebox
\usepackage{siunitx}
\usepackage[bookmarks=false]{hyperref}

\usepackage{xcolor}

\title{Multi-scale decomposition of sea surface height snapshots %into Balanced and Unbalanced components 
using machine learning}

\author{
  Jingwen Lyu$^*$ \\
  Applied Physics and Applied Mathematics \\
  Columbia University \\
  New York, USA \\
  \texttt{jl6811@columbia.edu} \\
  \And
  Yue Wang$^*$ \\
  Applied Physics and Applied Mathematics \\
  Columbia University \\
  New York, USA \\
  \texttt{yw4236@columbia.edu} \\
  \AND
  Christian Pedersen \\
  Courant Institute \\ 
  New York University \\
  New York, USA\\
  \texttt{c.pedersen@nyu.edu} \\
  \And
  Spencer Jones \\
  Department of Oceanography \\
  Texas A$\&$M University\\
  Texas, USA \\
  \texttt{spencerjones@tamu.edu} \\
  \AND
  Dhruv Balwada \\
  Lamont Doherty Earth Observatory \\
  Columbia University \\
  New York, USA \\
  \texttt{dbalwada@ldeo.columbia.edu}
}

\begin{document}

\maketitle

\begin{abstract}
%  Decomposing the SSH field is relevant for (insert climate reasons)
%Upcoming observations from SWOT will produce observations of unprecedented resolution
%Traditional non-ML approaches fail because xyz, or are limited by xyz
%We explore ML approaches, and find “vanilla” ML approaches are limited at high resolution due to MSE limitations. We explore a range of loss modifications and data augmentations, and find ZCA whitening produces strong performance gains, particularly at high frequencies
Knowledge of ocean circulation is important for understanding and predicting weather and climate, and managing the blue economy. This circulation can be estimated through Sea Surface Height (SSH) observations, but requires decomposing the SSH into contributions from balanced and unbalanced motions (BMs and UBMs). 
%This is especially pertinent for the SSH measured by the novel SWOT satellite, which promises to improve spatial resolution of the estimated flow by an order of magnitude. 
This decomposition is particularly pertinent for the novel SWOT satellite, which measures SSH at an unprecedented spatial resolution. 
Specifically, the requirement, and the goal of this work, is to decompose instantaneous SSH into BMs and UBMs.   
%Ocean circulation plays a critical role in Earth's climate since the ocean stores large amount of tracers, such as heat and carbon. 
%This circulation, or flow, can be estimated using satellite altimetry, which measures Sea Surface Height (SSH). However, estimating flow from SSH presents several challenges, as SSH is not directly or trivially related to the flow field.
%At a fundamental level, oceanic flow can be decomposed into balanced motions (BM) and unbalanced motions (UBM), each playing distinct roles in ocean dynamics, transport, and energetics. For instance, the majority of tracer transport is driven by BM, making it essential to accurately estimate BM to quantify the tracer uptake in the ocean.
%Traditional altimeters, in use since the early 1990s, measure along a 1D track and can resolve spatial scales as small as 100 km, where BM are dominant. However, the recently launched Surface Water and Ocean Topography (SWOT) satellite offers an unprecedented spatial resolution of 5-10 km and maps the ocean surface in 2D on a 21-day repeat cycle \citep{fu2024surface}. At these finer spatial resolutions, BM and UBM co-exist, offering new insights into ocean circulation. However, the low temporal resolution (21 days) does not allow the use of traditional temporal harmonic analysis methods for UBM-BM separation, and approaches that can work with a single SSH snapshot need to be developed.  
While a few studies using deep learning (DL) approaches have shown promise in framing this decomposition as an image-to-image translation task, these models struggle to work well across a wide range of spatial scales 
%and usually do not capture the finer-scale information necessary for precise separation. 
and require extensive training data, which is scarce in this domain. These challenges are not unique to our task, and pervade many problems requiring multi-scale fidelity.
%We introduce a novel DL method using the U-Net architecture to filter UBM from SSH snapshots, 
We show that these challenges can be addressed by using zero-phase component analysis (ZCA) whitening and data augmentation; making this a viable option for SSH decomposition across scales. %Our approach significantly improves skill across a wide range of scales and mitigates data scarcity. 
%, improving the accuracy of BM-UBM separation for high-resolution SSH. 
\end{abstract}

\section{Introduction}
% Intro paragraphs:
% Oceans are important for climate, and ocean circulation must be well estimated.
% How SSH is related to ocean circulation, and that ocean flow is composed of different types of motions. 
% 
Oceans play a critical role in Earth's climate and weather, as they store and circulate large quantities of heat, carbon, and other biogeochemically relevant tracers \citep{cronin2019air}\citep{dong2014global}. 
%Since the start of the industrial revolution, the oceans have absorbed over 25\% of the anthropogenic carbon and over 90\% of the excess heat that has accumulated in the climate system. 
%Also, the ocean and the atmosphere are a coupled system, which develops natural climate variability, impacting our lives through daily and weekly weather, modulated seasonal cycles, multi-year El-Nino and La-Nina oscillations, and so on \citep{clayson2023new, cronin2019air}. 
%Many of these impacts are shaped by the ocean circulation or flow of water in the ocean, which is composed of many different physical processes and varies across a wide range of spatial and temporal scales \citep{talley2011descriptive}. 
% This circulation is composed of many different physical processes and varies across a wide range of spatial and temporal scales \citep{talley2011descriptive},
%Knowledge of oceanic flow is also relevant for other reasons, like optimizing shipping routes, management of fisheries and marine sanctuaries, and planning and development of the ocean energy sector.
This circulation is composed of many different physical processes across a wide range of scales \citep{talley2011descriptive}. Inferring circulation patterns plays a vital role in understanding climate variability, as well as management of vital sectors like shipping, fisheries, ocean energy, and marine carbon dioxide removal.
% and its knowledge is not only relevant for understanding the climate variability, but also for management of sectors like shipping, fisheries, ocean energy, and marine carbon dioxide removal.
Currently the only way to observationally estimate this flow globally and across a wide range of scales is through satellite altimetry, which measures sea surface height (SSH). 
SSH corresponds to the pressure at the surface of the ocean, and is related to the surface flow. 
However, the relationship is not simple, and depends on the nature of the processes contributing to the flow.
%However, the relationship is not direct or trivial, and depends on the nature and scales of the processes contributing to the flow. 
These processes can be broadly divided into two categories, balanced motions (BM, slowly evolving) and unbalanced motions (UBM, rapidly evolving) \citep{torres2018partitioning, klein2019ocean}. 
%BMs are composed of slowly evolving flows like ocean gyres, ocean eddies, and boundary currents, while UBMs are composed of rapidly varying flows like tides, inertial oscillations and inertia-gravity waves \citep{torres2018partitioning, klein2019ocean}. 
%Along with this distinction in temporal scales, generally UBMs are associated with smaller spatial scales ($\sim 50$ km or smaller), while BMs are dominant on larger spatial scales. Also,
%BMs and UBMs have distinct relationships between the SSH and surface flows, and 
It is essential to decompose the SSH into the contributions from BMs and UBMs before attempting to estimate flow from SSH, particularly when using high spatial resolution SSH data. 
Theoretically, BMs and UBMs can be decomposed by applying temporal or spatio-temporal filters to data \citep{jones2023using}, but this conventional approach requires data with a temporal resolution on the order of minutes to a few hours.

%Traditional satellite altimeters, in use since the early 1990s, measure SSH along a 1D track. This 1D track data is mapped to 2D when multiple satellite passes are available, and these mapped data products resolves spatial scales of O(100)km and temporal scales of O(10) days. At these resolved scales BM are dominant, and the flow is estimated using the geostrophic relationship, which estimates the flow by using the spatial gradient of the SSH \citep{vallis2017atmospheric}. 
Traditional satellite altimeters are used to produce maps of SSH with spatial and temporal resolution of O(100)km and O(10) days. At these scales BMs are dominant, and no decomposition has been necessary. 
However, the recently launched Surface Water and Ocean Topography (SWOT) satellite provides data in a %is causing a paradigm shift, by measuring SSH along a 
2D swath at an unprecedented spatial resolution of 5-10 km \citep{fu2024surface, chelton2024post}, where BMs and UBMs have comparable magnitudes. %At this resolution, BM and UBM co-exist, potentially offering new insights into ocean circulation. 
However, the SWOT satellite only passes over a region once every 21 days, which does not allow the use of conventional methods for UBM-BM separation that require data at high temporal resolution. \textit{Thus, decomposition approaches that can work with only instantaneous SSH observations (snapshots) need to be developed, and this is the focus of our study.} 

With the rapid advancement of machine learning techniques, SSH decomposition may be approached as an Image-to-Image translation problem. %State-of-the-art U-Net architectures have demonstrated promising results high resolution image-to-image translation tasks \citep{laxman2022efficient}. 
A few studies have now used convolutional neural networks (CNNs) to decompose the BMs and UBMs \citep{lguensat2020filtering, wang2022deep, gao2024deep}, with some success. 
However, significant challenges persist due to the multi-scale nature of these fields, implying that there is important detail in the SSH fields across a large range of scales. Often the amount of variance or signal present at different scales can differ by orders of magnitudes, as exemplified by the power law decay of variance across wavenumbers in power spectral density (PSD) plots \citep{vallis2017atmospheric}. 
%. The power spectral densities (PSD) of BM and UBM SSH contribution decay as a power law relationship in wavenumber (often with a power law exponent varying between -5 to -2, depending on the dynamics of the flow). 

Pixel-wise mean squared error (MSE) loss fails to effectively capture patterns across the full range of scales for these multi-scale fields, because this choice of loss tends to prioritize learning the scales with the dominant signal \citep{rahaman2019on}. 
To address this issue, \citep{lguensat2020filtering} and \citep{gao2024deep} augmented their loss function with gradients of of the predicted field, which allowed the loss function to also focus on small scale information. 
While this approach improves the model's performance at smaller scales, it introduces two significant drawbacks: (1) the method requires careful tuning of the gradient loss weight, and this weight likely needs to be adjusted for every new training dataset with different intensities of BMs and UBMs; (2) the gradient loss can force the model to focus on high-wavenumber information, and with insufficient training data this can easily lead to overfitting in the highest-wavenumber range. 

To address the challenge of working with multi-scale data, we propose implementing ZCA transformation to whiten the UBM prior to processing. This approach offers several advantages: (1) It enhances  information across a range of scales, thereby reducing the necessity for gradient loss. (2) It maintains compatibility with pixel-wise Mean Squared Error (MSE) calculations and mitigates the risk of overfitting at extremely high frequencies. (3) ZCA transformation also approximates the full covariance matrix of samples to an identity matrix \citep{kessy2018optimal}, which effectively reduces correlation between samples, leading to improved training stability and computational efficiency.

\begin{figure}[htbp]
    \centering
    \includegraphics[width=\textwidth]{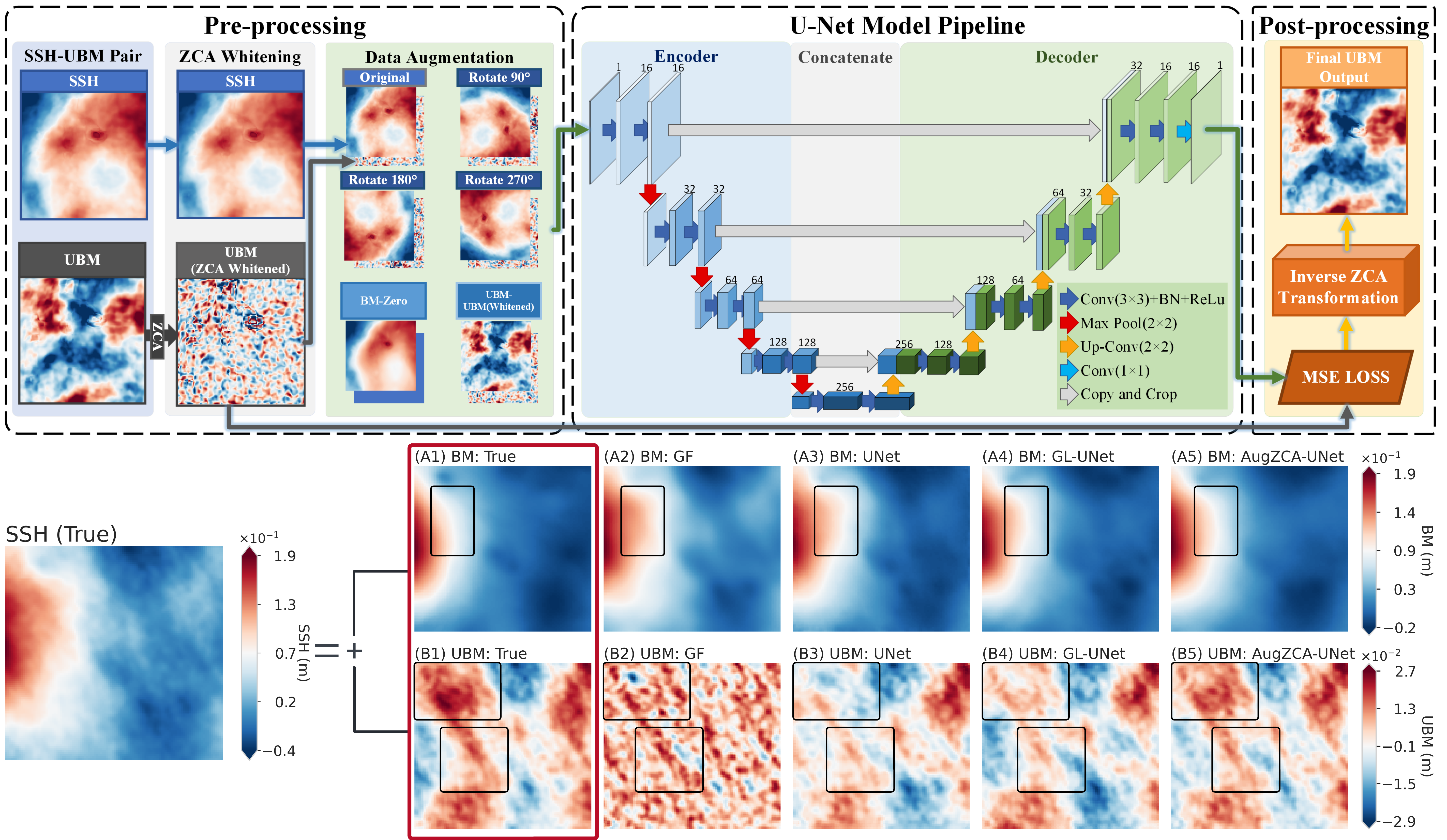}
    \label{fig:predictions_best}
    \caption{Top: Schematic representing model pipeline and architecture. Bottom: Visualization of UBM and BM for our best performing test sample using AugZCA-UNet. A1-A5: BM; B1-B5: UBM. Black squares are a visual aid.}
\end{figure}

\section{Methodology}
Our task is to decompose the total SSH ($\eta$) into contributions from BMs ($\eta_{\text{B}}$) and UBMs ($\eta_{\text{UB}}$), where $\eta=\eta_{\text{B}} + \eta_{\text{UB}}$. We achieve this by training a ML model that can predict the UBM ($\hat{\eta}_{\text{UB}}$) for a given input $\eta$. The prediction for BM is consequently defined as $\hat{\eta}_{\text{B}} = \eta - \hat{\eta}_{\text{UB}}$. %We describe below how the data for this task ($\eta$, $\eta_{\text{UB}}$ pairs) is procured and processed, and in the next section we describe the different ML models we tested. 

%\subsection{Data}
\textbf{Data:}
We utilize data from a high-resolution global ocean simulation, focusing on the Agulhas retroflection region \citep{jones2023using}. The dataset comprises 70 daily snapshots of total SSH, BM, and UBM from September to November 2011. This region presents challenges in distinguishing BM and UBM due to their amplitude disparity, diverse regional patterns, and the significance of UBM in gradient fields. Detailed information about the dataset and its processing is provided in Appendix ~\ref{Appendix:Data}. 

To mitigate data scarcity, we employed two data augmentation techniques: (1) rotational augmentation (90°, 180°, and 270°) for promoting rotational invariance, and (2) synthetic sample generation of purely BM or UBM flows by adding pairs ($\eta_{\text{B}}$, $\eta_{\text{UB}}=0$) and ($\eta_{\text{UB}}$, $\eta_{\text{UB}}$). As a final and crucial preprocessing step, we applied \textit{ZCA whitening} to the output, $\eta_{\text{UB}}$. This technique, detailed in Appendix ~\ref{Appendix:ZCA}, represents a key contribution of our methodology, as it enhances the model's ability to detect and process both large-scale and fine-scale features in the data.

\textbf{ML Models:}
To assess the effectiveness of our ML-based techniques and the impact of data augmentation and ZCA whitening, we systematically compared the following approaches: (1) Gaussian Filter (GF)—a non-ML baseline detailed in Appendix \ref{Appendix:Gaussian Filter}; (2) UNet—an ML baseline without preprocessing or additional gradient loss terms, described in Appendix \ref{Appendix:UNet}; (3) UNet with gradient loss (GL-UNet)—incorporating gradient loss terms with varying weights ($\alpha = 0.5, 1, 5, 10$), with details provided in Appendix \ref{Appendix:Gradient Loss}; (4) UNet with data agumentation (Aug-UNet)—employing rotation and synthetic data generation techniques; (5) UNet coupled with ZCA whitening (ZCA-UNet); and (6) UNet coupled with both data augmentation and ZCA whitening (AugZCA-UNet). Notably, all ML models share the same UNet configuration as the `raw' UNet, and models employing ZCA did not include gradient loss terms, allowing us to isolate the effects of each technique.

% \textbf{Evaluation Metrics}
Model performance was rigorously evaluated using two complementary metrics: (i) Pixel-wise absolute error distribution of $\hat{\eta}_{\text{B}}$ and $|\nabla \hat{\eta}_{\text{B}}|$, which assesses local prediction accuracy across the spatial domain, and (ii) Power Spectral Density (PSD) of $\hat{\eta}_{\text{B}}$ and $\hat{\eta}_{\text{UB}}$, which evaluates the model's ability to capture multi-scale characteristics across spatial scales, providing insight into the multi-scale fidelity of the predictions.

\section{Results}
A visual comparison of SSH decomposition methods into BM and UBM reveals that all ML approaches show skill in this task (Figure \ref{fig:predictions_best}, additional samples in Figure \ref{fig:predictions_median_worst}). The ML models effectively reduce fine-scale UBM features in the gradient field, which is crucial for flow estimation methods requiring accurate BM gradients. Unlike the GF, ML based models avoid over-smoothing the predicted BM and its gradients.

Quantitative evaluation using pixel-wise absolute error shows AugZCA-UNet consistently outperforms other methods for both BM and its gradient across all statistical measures (Table \ref{tab:combined-pixel-wise-error-dist-simple}, Appendix ~\ref{Appendix:Visualization}). Conversely, GF consistently performs the worst, indicating ML approaches can surpass traditional filters. GL-UNet's performance varies parabolically with gradient weight, peaking at 1 and declining at lower or higher values, highlighting the need for precise gradient weight tuning. ZCA-UNet and AugZCA-UNet, requiring no gradient weight tuning, demonstrate superior performance due to their architectural adjustments. Data augmentation techniques also consistently enhance model performance.

% A quantitative evaluation using pixel-wise absolute error shows that
% AugZCA-UNet consistently yields superior performance across all statistical measures for both the BM and its gradient (Table \ref{tab:combined-pixel-wise-error-dist-simple} and Appendix ~\ref{Appendix:Visualization}). In contrast, GF consistently demonstrates the poorest performance across all these statistical measures for both the BM and its gradient, suggesting that ML based approaches can surpass traditional filters. %This suggests that a non-ML approach struggles to accomplish such sophisticated tasks.
% For GL-UNet, the performance changes roughly like a parabola as a function of the gradient weight, performing the best 
% %the weights of the gradient loss act like a parabola in terms of performance, optimizing 
% at a weight of 1 but deteriorating at lower or higher values for our case, indicating the necessity for precise tuning of this parameter. Unlike GL-UNet, both ZCA-UNet and AugZCA-UNet do not require gradient weight tuning and demonstrate superior performance, suggesting inherent advantages in their architectural adjustments. Additionally, data augmentation techniques consistently improve model performance.
\begin{table}[h]
  \caption{Pixel-wise Absolute Error Distribution Measures for $\hat{\eta}_{\text{B}}$ ($\times 10^{-2}$) and $|\nabla \hat{\eta}_{\text{B}}|$ ($\times 10^{-3}$)}
  \label{tab:combined-pixel-wise-error-dist-simple}
  \centering
  %\small
  \scalebox{0.76}{
    \begin{tabular}{llccccccccc}
      \toprule
      \multirow{2}{*}{} & \multirow{2}{*}{Measures} & \multirow{2}{*}{GF} & \multicolumn{5}{c}{GL-UNet} & \multirow{2}{*}{Aug-UNet} & \multirow{2}{*}{ZCA-UNet} & \multirow{2}{*}{AugZCA-UNet}\\
      \cmidrule(lr){4-8}
      & & & $\alpha=0$ & $\alpha=0.5$ & $\alpha=1$ & $\alpha=5$ & $\alpha=10$ & & & \\
      \midrule
      \multirow{2}{*}{$\hat{\eta}_{\text{B}}$} 
      & Median & $0.406$ & $0.404$ & $0.420$ & $0.397$ & $0.398$ & $0.406$ & $0.384$ & $0.393$ & 
      $\textbf{0.372}$ \\
      & P95 (95\%) & $1.68$ & $1.52$ & $1.49$ & $1.48$ & $1.51$ & $1.53$ & $1.50$ & $1.52$ & $\textbf{1.40}$ \\
      \midrule
      \multirow{2}{*}{\shortstack{{\large $|\nabla \hat{\eta}_{\text{B}}|$}}}
      & Median & $0.505$ & $0.464$ & $0.447$ & $0.450$ & $0.450$ & $0.453$ & $0.460$ & $0.450$ & $\textbf{0.427}$ \\
      & P95 (95\%) & $2.20$ & $1.80$ & $1.70$ & $1.72$ & $1.74$ & $1.75$ & $1.78$ & $1.71$ & $\textbf{1.64}$ \\
      \bottomrule
    \end{tabular}
  } % end of scalebox
\end{table}
For many oceanographic and climate science applications, it is important that the ML models perform well across a wide range scales. To assess the prediction skill across spatial scales, we consider the wavenumber PSD of the predictions and the $R^2$ (defined in Appendix ~\ref{Appendix:R Ratio}), as shown in Figure \ref{fig:psds}. A more rigorous quantitative comparison is presented in Appendix ~\ref{Appendix:Power Spectral Density}. GF, constrained by its fixed filter length scale, only works well on scales where BM or UBM are dominant, and is unable to perform at wavenumbers where they have comparable magnitues or in cases where we are interested in predicting the sub-dominant signal. In contrast, the UNet models exhibit varying degrees of success across the wavenumber spectrum. At low wavenumbers (below $10^{-1}$ cpkm), all UNet models achieve relatively high accuracy in predicting the true PSD. %This performance can be attributed to UNet's proficiency in extracting large-scale features, further bolstered by the stronger signal presence at these scales.
However, at higher wavenumbers the performance characteristics of DL models diverge. 
The GL-UNet can be tuned to outperform a raw UNet, but often suffers from noisy prediction at the highest wavenumber that can severely degrade the skill in predicting gradient fields.
%enhancing precision within wavenumbers ranging from $10^{-1}$ to $6 \times 10^{-1}$ cpkm, corresponding to wavelengths of 10 km to 1.67 km. However, beyond $6 \times 10^{-1}$ cpkm, a marked PSD spike appears in the GL-UNet results, diverging notably from the expected spectrum. 
In contrast the ZCA-UNet and AugZCA-UNet consistently exhibit superior performance across the full wavenumber spectrum, and are not plagued by noise at the highest wavenumbers.  

%, including the larger wavenumber range where PSD is nearly five orders of magnitude smaller than at small wavenumbers, demonstrating a robustness that surpasses GF, UNet, and GL-UNet models.
\begin{figure}[htbp]
    \centering\includegraphics[width=\textwidth]{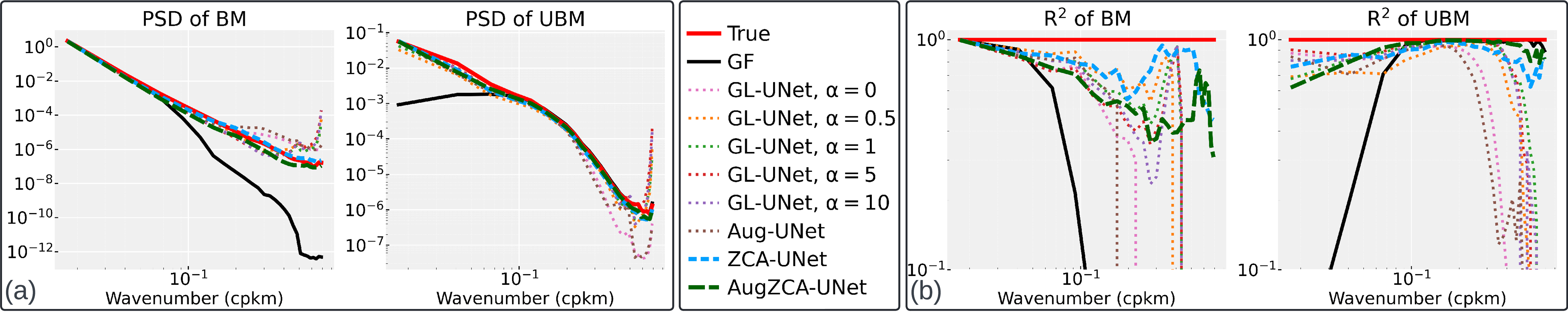}
    \caption{Comparison of PSD and $R^2$ across various models. \textbf{(a)} Mean PSD over test samples for UBM (left) and BM (right). \textbf{(b)} $R^2$ for UBM (left) and BM (right).}
    \label{fig:psds}
\end{figure}

%These results highlight the complex interplay between preprocessing methods and regularization techniques in spectral analysis. While gradient loss offers localized benefits in low PSD capture, its broader application requires careful tuning to avoid high-frequency distortions. In contrast, ZCA preprocessing emerges as a robust strategy, enhancing model performance across the entire wavenumber spectrum. This striking difference in outcomes underscores the critical role of preprocessing in addressing spectral analysis challenges. The superior performance of ZCA-based models warrants further investigation into their underlying mechanisms and potential synergies with other techniques.

\section{Discussion and Outlook}
Our novel AugZCA-UNet shows significant promise in skillfully decomposing SSH at finer scales with lower variance, while partially addressing data scarcity. It outperforms the previously proposed GL-UNet, demonstrating superior performance in both pixel-wise error (sensitive to dominant signals) and PSD accuracy (sensitive to multi-scale information). ZCA whitening is key to our model's success, which enhances the loss function's sensitivity to fine-scale variations that are otherwise overshadowed by dominant large-scale variations. This technique also has the potential to mitigate challenges across a broad spectrum of climate-related research where capturing multi-scale phenomena is crucial \citep{lai2024machine}.

While our work marks an important step forward in applying ML to multi-scale systems, several challenges must be addressed before our decomposition approach can be effectively applied to actual SSH data from SWOT.  
% Image size limitations 
Firstly, ZCA constrains the ML model to a specific image size, compromising the fully convolutional nature of UNets. Therefore, pragmatic processing choices will be necessary when working with datasets that do not naturally align with the image sizes used during model training or when making predictions over large domains. 
% Memory challenge 
Secondly, the memory cost of performing ZCA increases with image size, necessitating modifications to the standard ZCA algorithm for more efficient memory management in tasks requiring larger images. 
% Working with gappy data 
Thirdly, while our dataset included gappy data due to land and other numerical artifacts, we have not thoroughly assessed the model's performance near these gaps. This is particularly crucial for SWOT, where the satellite swath is 120 km wide but consists of two 50 km sections separated by a 20 km gap.  
% Availability of real obs data, instead of relying on high resolution simulations % Generalization to other regimes - other geography or seasons needs to be evaluated
 % In this study, we trained on a specific region of the ocean, during a particular time period, and using a high-resolution model. We do not expect the exact same model to work universally across diverse geographical regions, seasons, or real observations. 
Finally, the generalization of our model, particularly relative to spatial filters like GF, remains an open question. In future work, we plan to evaluate the model’s generalization properties and explore whether flow-dependent normalization \citep{beucler2024climate} or transfer learning approaches \citep{xiao2023reconstruction} can enhance its applicability.

\bibliographystyle{plainnat}
\bibliography{main}

\newpage

%\begin{center}
\section*{\centering APPENDIX}
%\end{center}
\appendix
%\renewcommand{\thefigure}{\thesection.\arabic{figure}}
% Reset figure counter and redefine figure numbering for SI
\setcounter{figure}{0}
\renewcommand{\thefigure}{App.\arabic{figure}}
\setcounter{table}{0}
\renewcommand{\thetable}{App.\arabic{table}}

\section{Data and Code Availability}
Data is currently stored openly on Zenodo, and both data and the code used in the work will be made publicly available after de-anonymization via a github repository.

\section{Comprehensive Overview of Dataset}
\label{Appendix:Data}
Ocean models solve the fundamental equations of fluid flow, primitive equations, under realistic boundary conditions and forcing. At high-resolutions these simulations can generate a statistically well matched representation of the real world \citep{arbic2022near, ansong2024surface}. Here we utilized the data generated by one of the highest resolution global ocean simulations that is currently available, referred to in the climate science community as LLC4320. While SSH observed by SWOT has not yet been decomposed into BMs and UBMs, we utilized a publicly available decomposed subset of the LLC4320 data \citep{jones2023using}. This was achieved using spatio-temporal filters, as the LLC4320 data is available at a high temporal resolution of 1 hour, making it directly amenable to conventional decomposition methods. The primary objective of our study is to assess the effectiveness of ML in performing this decomposition using only instantaneous/snapshot data, thereby addressing the temporal resolution limitations of SWOT satellite observations.

The spatial domain covers the Agulhas retroflection region (15\textdegree W to 29\textdegree E and 27\textdegree S to 57\textdegree S), located to the south and west of Africa. The temporal domain comprises 70 daily snapshots spanning from September 14, 2011 to November 22, 2011. We divided the dataset through temporal partitioning (days 0--60 for training, 61--65 for validation, and 66--70 for testing) and spatial tiling (400 non-overlapping 108 $\times$ 108 pixel patches from a 2160 $\times$ 2160 pixel domain), yielding 24,000 training images. The land regions (represented by NaN values) are masked when calculating the loss function, thereby reducing their influence on the model training.

In this spatiotemporal domain, accurately distinguishing between balanced and unbalanced components from snapshots presents three significant challenges:

\begin{enumerate}
    \item \textbf{Scale disparity:} UBM has a magnitude approximately one-tenth that of total SSH. This disparity hinders CNNs in detecting UBM features without preprocessing.
    
    \item \textbf{Diverse dynamics:} The dataset exhibits varied dynamical patterns across spatial regions, necessitating a robust scheme capable of isolating UBM across different oceanic dynamics.
    
    \item \textbf{Gradient field complexity:} Despite their small magnitude, the UBM gradient fields exhibit amplitudes comparable to both BM and SSH. This similarity hinders the application of typical methods, such as geostrophic balance, to estimate the flow.
\end{enumerate}

These challenges underscore the importance of effective UBM filtering for accurate ocean dynamics analysis.

\begin{figure}[htbp]
    \centering
    \includegraphics[width=0.95\textwidth]{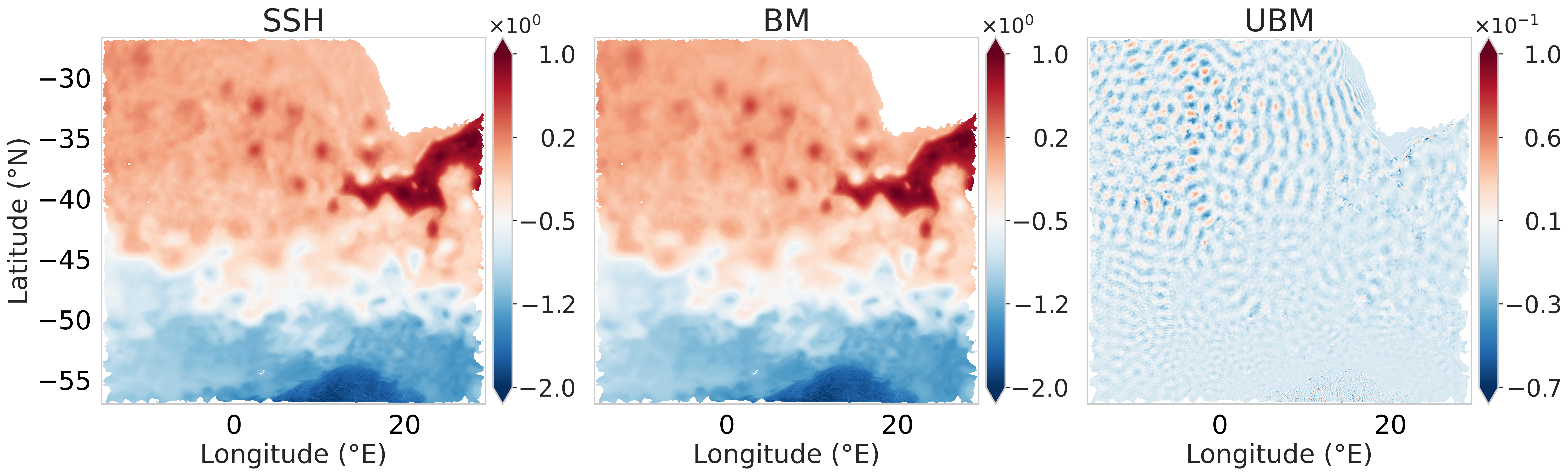}
    \caption{A snapshot of the whole region SSH, BM and UBM}
    \label{}
\end{figure}

\section{Zero-phase Component Analysis (ZCA)}
\label{Appendix:ZCA}
Let \( X_{\eta_{\text{UB}}} \in \mathbb{R}^{n \times d} \) represent the data matrix of UBM, where \( n \) is the number of training samples of UBM, and \( d = 108 \times 108 \) is the number of flattened features for a single UBM snapshot. The ZCA whitening process is implemented as follows:
\begin{enumerate}
    \item \textbf{Data Centering}: Compute the centered data matrix \( X_{\eta_{\text{UB}}}^c = X_{\eta_{\text{UB}}} - \mu_{\eta_{\text{UB}}} \) where \( \mu_{\eta_{\text{UB}}} \) is the mean vector computed across all training UBM samples.
    
    \item \textbf{Covariance Matrix Computation}: Calculate \( \Sigma = \frac{1}{n-1} (X_{\eta_{\text{UB}}}^c)^T X_{\eta_{\text{UB}}}^c \).
    
    \item \textbf{Eigendecomposition}: Decompose \( \Sigma = U \Lambda U^T \), where \( U \) is the matrix of eigenvectors and \( \Lambda \) is the diagonal matrix of eigenvalues.
    
    \item \textbf{Compute ZCA Matrix}: Calculate \( W_{\text{ZCA}} = U (\Lambda + \epsilon I)^{-1/2} U^T \) where \( I \) is the identity matrix and \( \epsilon \) is a small parameter added for numerical stability. This parameter controls the extent of whitening by slightly perturbing the eigenvalues. In our experiments, we set \( \epsilon = 10^{-5} \). We found that performance was not highly sensitive to variations in $\epsilon$, testing values from $10^{-3}$ to $10^{-7}$.
    
    \item \textbf{Apply Transformation}: Transform all UBM samples (training, validation, and test sets) using the same mean and ZCA matrix calculated from the training data: \( X_{\text{ZCA}} = X_{\eta_{\text{UB}}}^c W_{\text{ZCA}} \)
\end{enumerate}

In the frequency domain, the covariance matrix is intrinsically linked to the Power Spectral Density (PSD) according to the Wiener-Khinchin theorem, which states that the PSD is the Fourier transform of the autocorrelation function \citep{wiener1930generalized}. Whitening the data through ZCA transforms the full covariance matrix into an identity matrix, enforcing a uniform power distribution across all frequencies. This process effectively results in a flatter PSD post-whitening transformation.

\begin{figure}[htbp]
    \centering  \includegraphics[width=0.95\textwidth]{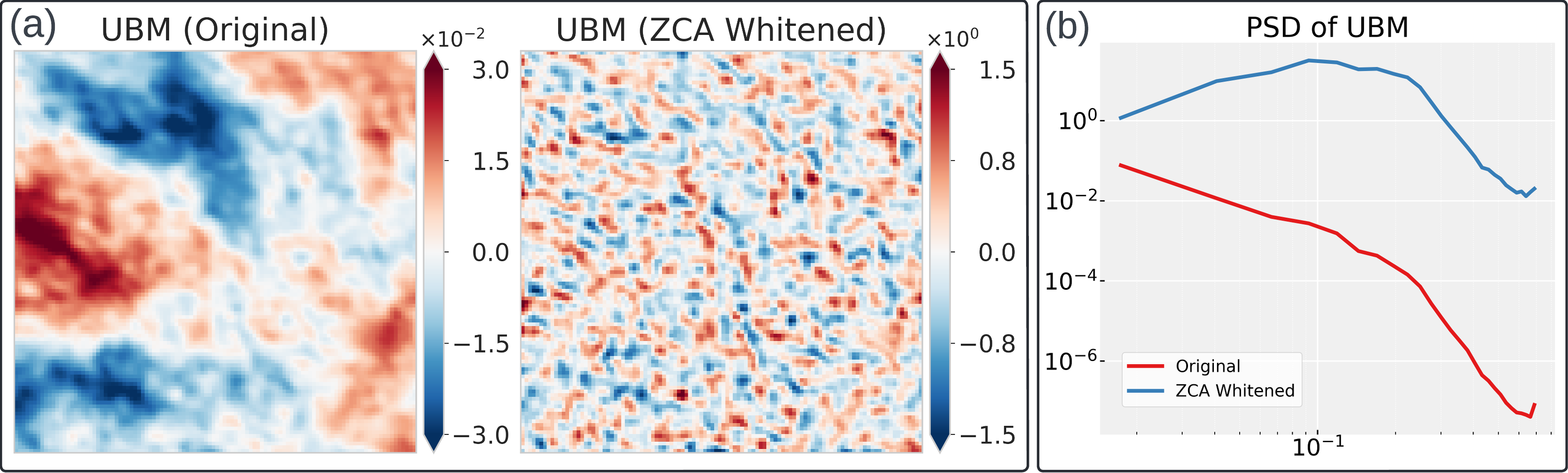}
    \caption{Effect of ZCA whitening on the original UBM snapshot (a) and its PSD (b)}
    \label{fig:zca}
\end{figure}

Figure \ref{fig:zca} illustrates an example of the impact of ZCA whitening on a UBM snapshot sample. The ZCA process notably enhances small-scale features, resulting in a more granular appearance in the spatial domain and a flatter spectrum in the frequency domain (Note that the whitened spectrum varies across about 3 orders of magnitude, while the original spectrum had a variation across about 7 orders of magnitude).

\section{UNet Architecture and Training Specifications}
\label{Appendix:UNet}
The UNet architecture employed in this study utilizes Rectified Linear Unit (ReLU) activation functions and the Adam optimizer. We implemented a dynamic learning rate schedule starting at an initial rate of $10^{-3}$. This rate is adjusted during training with a reduction factor of 0.8, which is applied when the validation loss plateaus for three consecutive epochs. To guard against overfitting, we incorporated an early stopping mechanism with a patience of 20 epochs. This approach allows the model to continue training as long as it shows improvement, but halts the process if no progress is observed over an extended period.

The complete network configuration is illustrated in Figure \ref{fig:predictions_best}. It's important to note that this configuration remained consistent across all UNet variants used in our experiments. This consistency was maintained to ensure a fair comparison between different approaches and to isolate the effects of our proposed modifications.

\section{Visualization of Prediction Outcomes}
\label{Appendix:Visualization}
\begin{figure}[htbp]
    \centering
    \includegraphics[width=0.95\textwidth]{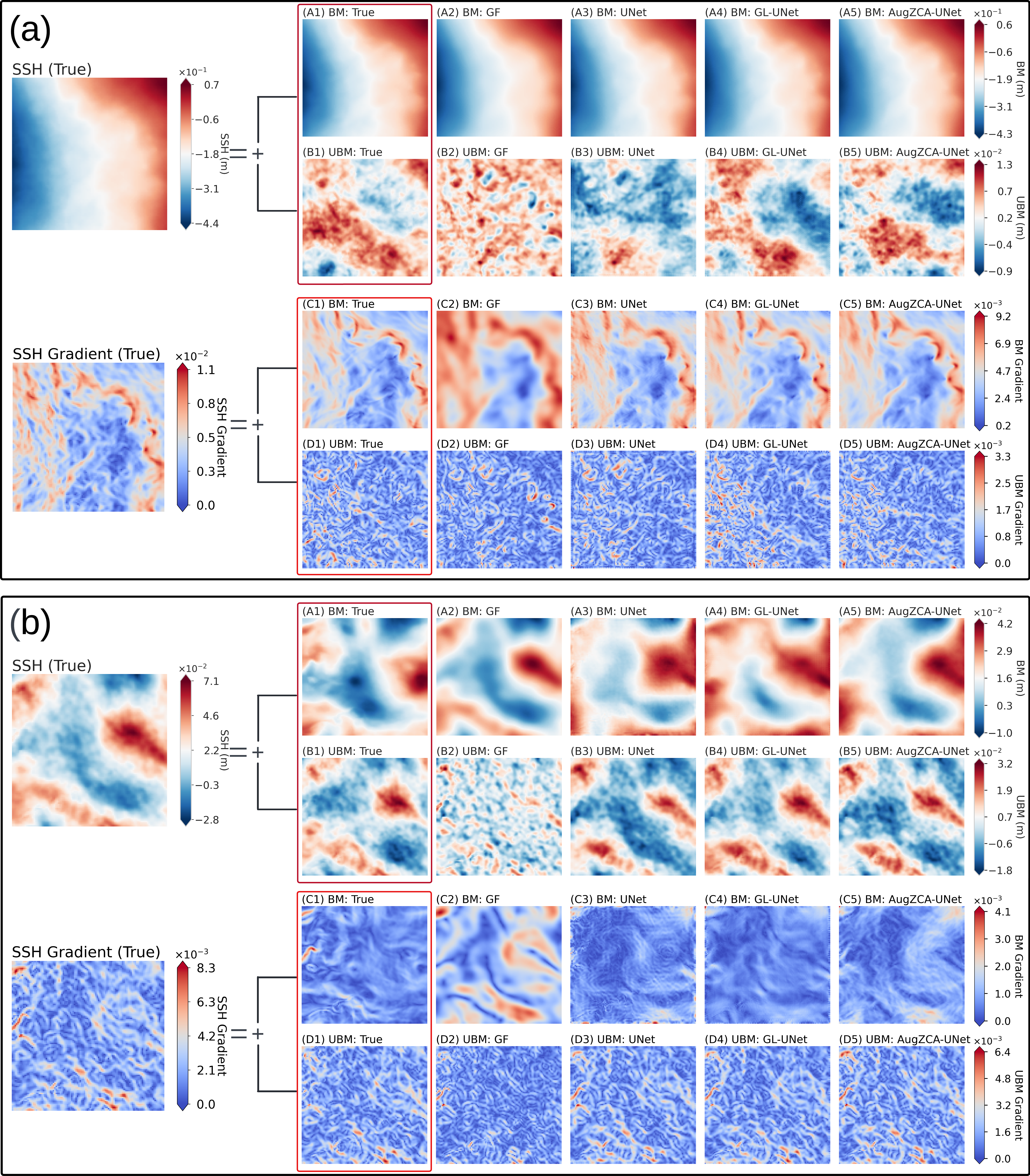}
    \caption{Visualization of two test samples based on correlation coefficients between AugZCA-UNet predicted and true BM gradients: (a) Sample with median correlation (0.936); (b) Sample with lowest correlation (0.232). A1-A5: BM; B1-B5: UBM; C1-C5: BM Gradient; D1-D5: UBM Gradient.}
    \label{fig:predictions_median_worst}
\end{figure}

\section{Gaussian Filter (GF)}
\label{Appendix:Gaussian Filter}
Gaussian filters are fundamental tools in signal and image processing, prized for their isotropic and radially symmetric properties. They effectively attenuate high-frequency components while preserving low-frequency information, and have been used a lot in the filtering of gridded geophysical data \citep{grooms2021diffusionbased}. The filter is applied to an image or signal $f(x)$ through convolution, represented by the following integral:

\[(f * G)(x) = \int f(u) \cdot G(x-u) \, du\]

The GF approach, which acts as a spatial filter that effectively truncates the SSH power spectra at a specific wavenumber, effectively removing high-frequency components typically associated with UBM. 

The scale of the filter is inferred from the power spectra of the total SSH, which often shows a change in spectral properties at scales where UBM start to dominate over BM \citep{torres2019diagnosing}. Here we used a single scale of 24km for the entire domain considered.

\section{Error Distribution}
\label{Appendix:Error Distribution}
Table \ref{tab:pixel-wise-error} presents the pixel-wise absolute error distribution measures for $\hat{\eta}_{\text{B}}$ and $|\nabla \hat{\eta}_{\text{B}}|$, which are computed as follows:

\[
\epsilon_{\eta_{\text{B}}} = |\hat{\eta}_{\text{B}} - \eta_{\text{B}}|
\]
\[
\epsilon_{\nabla\eta_{\text{B}}} = \big| |\nabla\hat{\eta}_{\text{B}}| - |\nabla\eta_{\text{B}}| \big|
\]
\begin{table}[ht]
\caption{Pixel-wise Absolute Error Distribution Measures for $\hat{\eta}_{\text{B}}$ ($\times 10^{-2}$) and $|\nabla \hat{\eta}_{\text{B}}|$ ($\times 10^{-3}$)}
\label{tab:pixel-wise-error}
\centering
\scalebox{0.8}{
\begin{tabular}{@{}llccccccccc@{}}
\toprule
& \multirow{2}{*}{Measures} & \multirow{2}{*}{GF} & \multicolumn{5}{c}{GL-UNet} & \multirow{2}{*}{Aug-UNet} & \multirow{2}{*}{ZCA-UNet} & \multirow{2}{*}{AugZCA-UNet} \\
\cmidrule(lr){4-8}
& & & $\alpha=0$ & $\alpha=0.5$ & $\alpha=1$ & $\alpha=5$ & $\alpha=10$ & & & \\
\midrule
\multirow{5}{*}{$\hat{\eta}_{\text{B}}$} 
& Mean    & 0.573 & 0.542 & 0.546 & 0.530 & 0.536 & 0.546 & 0.524 & 0.533 & \textbf{0.498} \\
& Median  & 0.406 & 0.404 & 0.420 & 0.397 & 0.398 & 0.406 & 0.384 & 0.393 & \textbf{0.372} \\
& Q1 (25\%)& 0.182 & 0.182 & 0.198 & 0.178 & 0.179 & 0.182 & 0.173 & 0.177 & \textbf{0.170} \\
& Q3 (75\%)& 0.774 & 0.757 & 0.757 & 0.744 & 0.749 & 0.765 & 0.724 & 0.740 & \textbf{0.687} \\
& P95 (95\%)& 1.68 & 1.52 & 1.49 & 1.48 & 1.51 & 1.53 & 1.50 & 1.52 & \textbf{1.40} \\
\midrule
\multirow{5}{*}{$|\nabla \hat{\eta}_{\text{B}}|$}
& Mean    & 0.795 & 0.656 & 0.626 & 0.636 & 0.640 & 0.642 & 0.651 & 0.634 & \textbf{0.604} \\
& Median  & 0.505 & 0.464 & 0.447 & 0.450 & 0.450 & 0.453 & 0.460 & 0.450 & \textbf{0.427} \\
& Q1 (25\%)& 0.229 & 0.213 & 0.205 & 0.207 & 0.206 & 0.208 & 0.211 & 0.207 & \textbf{0.196} \\
& Q3 (75\%)& 0.947 & 0.849 & 0.811 & 0.819 & 0.824 & 0.828 & 0.841 & 0.816 & \textbf{0.775} \\
& P95 (95\%)& 2.20 & 1.80 & 1.70 & 1.72 & 1.74 & 1.75 & 1.78 & 1.71 & \textbf{1.64} \\
\bottomrule
\end{tabular}
}
\end{table}

\section{Power Spectral Density (PSD) Analysis}
\label{Appendix:Power Spectral Density}
Table \ref{tab:log-rmse-psd-ubm-bm} presents the log-scale root mean square error (RMSE) of the Power Spectral Density (PSD), denoted as $\epsilon_{\text{PSD}}$, calculated across wavenumbers for the mean PSD of test samples:

\[
\epsilon_{\text{PSD}} = \sqrt{\frac{1}{|\mathcal{K}|} \sum_{k\in\mathcal{K}} (\log \mathbb{E}[S_{\text{ref}}(k)] - \log \mathbb{E}[S_{\hat{\eta}}(k)])^2}
\]

Here, $S_{\text{ref}}(k)$ and $S_{\hat{\eta}}(k)$ represent the true and predicted PSDs at wavenumber $k$, respectively, while $\mathbb{E}[\cdot]$ denotes the mean over test samples. This logarithmic approach allows for meaningful comparisons across the wide range of PSD magnitudes typical in spectral analysis.

To evaluate ML model performance at different scales, we calculate $\epsilon_{\text{PSD}}$ separately for small (< 0.2) and large (> 0.35) wavenumbers, corresponding to large-scale and small-scale features, respectively.

% Table \ref{tab:log-rmse-psd-ubm-bm} presents the log-scale root mean square error (RMSE) over wavenumbers of the mean PSD across test samples, denoted as ${\LARGE \epsilon_{\text{PSD}}}$. It is calculated as:
% \[
% \epsilon_{\text{PSD}} = \sqrt{\frac{1}{|\mathcal{K}|} \sum_{k\in\mathcal{K}} (\log \mathbb{E}[S_{\text{ref}}(k)] - \log \mathbb{E}[S_{\hat{\eta}}(k)])^2}
% \]
% where $S_{\text{ref}}(k)$ and $S_{\hat{\eta}}(k)$ represent the true and predicted PSDs of the test samples, respectively, at wavenumber $k$. $\mathbb{E}[\cdot]$ denotes the expectation (mean) over test samples. The use of logarithms enables comparison of predictions across a wide range of PSD magnitudes. We calculate this metric separately for small (< 0.2) and large (> 0.35) wavenumbers to evaluate ML model performance in capturing both large-scale and small-scale features, respectively.

\begin{table}[ht]
  \caption{Log-Scale RMSE of PSD Across Wavenumber Ranges for ML Models}
  \label{tab:log-rmse-psd-ubm-bm}
  \centering
  \scalebox{0.96}{
  \begin{tabular}{@{}lccccccccc@{}}
    \toprule
    & \multicolumn{5}{c}{GL-UNet} & \multirow{2}{*}{Aug-UNet} & \multirow{2}{*}{ZCA-UNet} & \multirow{2}{*}{AugZCA-UNet} \\
    \cmidrule(lr){2-6}
    & $\alpha=0$ & $\alpha=0.5$ & $\alpha=1$ & $\alpha=5$ & $\alpha=10$ & & & \\
    \midrule
    \multicolumn{9}{@{}l}{\textit{Small wavenumbers (< 0.2)}} \\
    UBM & 0.34 & 0.22 & 0.28 & \textbf{0.21} & 0.25 & 0.38 & 0.29 & 0.25 \\
    BM  & 0.41 & 0.67 & 0.61 & 0.69 & 0.58 & 0.39 & \textbf{0.27} & 0.71 \\
    \midrule
    \multicolumn{9}{@{}l}{\textit{Large wavenumbers (> 0.35)}} \\
    UBM & 2.59 & 1.11 & 0.72 & 1.04 & 0.94 & 2.18 & 0.41 & \textbf{0.29} \\
    BM  & 2.00 & 2.48 & 1.82 & 2.41 & 2.11 & 2.20 & 0.43 & \textbf{0.76} \\
    \bottomrule
  \end{tabular}
  }
\end{table}

\section{Gradient Loss}
\label{Appendix:Gradient Loss}
The loss function for the GL-UNet models involves the mean square error (MSE) of the UBM and its gradient:
\[
\mathcal{L}(\eta_{\text{UB}}, \tilde{\eta}_{\text{UB}}) = \|\eta_{\text{UB}} - \tilde{\eta}_{\text{UB}}\|^2 + \alpha \|\nabla \eta_{\text{UB}} - \nabla \tilde{\eta}_{\text{UB}}\|^2
\]
where \(\eta_{\text{UB}}\) is the ground truth UBM from training data set, \(\tilde{\eta}_{\text{UB}}\) is the prediction from the network, and \(\alpha\) is the weighting parameter. The choice of this loss function takes into account both the accuracy of the prediction and the preservation of high-frequency information for the UBM field.

\section{Scale-wise $R^{2}$}
\label{Appendix:R Ratio}
Figure~\ref{fig:psds} displays the averaged Power Spectral Density (PSD) for both UBM and BM in the first row, and the $R^{2}$ in the second row. The $R^2$ is calculated as follows:
\[
R^2 = 1 - \frac{\text{MSE}(\text{Predicted PSDs})}{\text{Var}(\text{True PSDs})} = 1 - \frac{E[(S_{\text{ref}}-S_f)^2]}{E[S_{\text{ref}}^2]}
\]
where $S_{\text{ref}}$ and $S_{f}$ are the true PSDs and predicted PSDs respectively. This formula reflects the relative quality of the estimation as a function of the wavenumber. Values of $R^2$ close to 1 indicate a perfect estimation.

\end{document}